\documentclass[prd,aps,twocolumn,showpacs,superscriptaddress,amsmath,floatfix,amssymb]{revtex4}
\usepackage{graphicx}

\newcommand{\eq}[1]{(\ref{#1})}

\newcommand{\beqn}{\begin{eqnarray}}
\newcommand{\eeqn}{\end{eqnarray}}

\newcommand{\lr}[1]{ \left( #1 \right) }
\newcommand{\lrs}[1]{ \left[ #1 \right] }

\newcommand{\vev}[1]{ \langle \, #1 \, \rangle }

\newcommand{\tr}{ {\rm Tr} \, }

\newcommand{\cO}{ {\cal O}}

\newcommand{\expa}[1]{ \exp{\left( #1 \right)} }

\begin{document}
\sloppy

\title{Quark electric dipole moment induced by magnetic field}

\author{P.V. Buividovich}
\affiliation{JINR, Dubna, Moscow region, 141980 Russia}
\affiliation{ITEP, B. Cheremushkinskaya 25, Moscow, 117218 Russia}

\author{M.N. Chernodub}
\affiliation{CNRS, LMPT, F\'ed\'eration Denis Poisson, Universit\'e de Tours, 37200 France}
\affiliation{DMPA, University of Gent, Krijgslaan 281, S9, B-9000 Gent, Belgium}
\affiliation{ITEP, B. Cheremushkinskaya 25, Moscow, 117218 Russia}

\author{E.V. Luschevskaya}
\affiliation{ITEP, B. Cheremushkinskaya 25, Moscow, 117218 Russia}

\author{M.I. Polikarpov}
\affiliation{ITEP, B. Cheremushkinskaya 25, Moscow, 117218 Russia}

\date{\today}
\begin{abstract}
We show numerically that quarks develop an electric dipole moment in the direction of a sufficiently intense
magnetic field due to local fluctuations of topological charge.  This anomalous $CP$-odd effect is a spin analogue of
the chiral magnetic effect in QCD.
\end{abstract}

\maketitle

\section{Introduction}

Fluctuations of topological charge is an important feature of the QCD vacuum which is reflected in the mass spectrum of hadrons,
in the breaking of chiral symmetry, and other properties.
Recently it was suggested that the topological fluctuations may be directly observed in hot QCD matter
in the presence of a very intense (hadron-scale) external magnetic field via the so called ``chiral magnetic effect'' (CME)~\cite{Kharzeev:08:1}.
The fluctuations of the topological charge lead to a local imbalance between the left- and right-handed
light quarks. In the presence of an intense magnetic field the quarks move along the field,
and the chiral imbalance creates a net {\it electric} current along the direction of the magnetic field.
As a result, this $CP$-odd  effect should lead to a separation of the electric charges along the field,
which potentially  may be observed in experiment.

The strong magnetic fields of the order of QCD scale may be created in noncentral heavy-ion collisions due to relative
motion of electrically charged ions in an initial state and the products of the collision in a final state~\cite{Kharzeev:08:1,Skokov}.
The strong fields are perpendicular to the collision plane, so that the CME should be realized on an event-by-event
basis in a form of an asymmetry of electrically charged particles emitted below and above the reaction plane.
There is a recent experimental evidence reported by the STAR collaboration that this scenario works at RHIC~\cite{Selyuzhenkov:06:1}.
Recently, the CME was also found in numerical lattice simulations~\cite{ref:CME} (a brief review of
lattice QCD simulations in strong magnetic fields can be found in Ref.~\cite{ref:Review}).

In this paper we show that the topological fluctuations in the background of the intense magnetic
field lead to appearance of a quark electric dipole moment. This effect is a spin analogue of
the CME. Indeed, the CME induces an electric dipole moment due to the  spatial (nonlocal) separation of
the electric charges of quarks while the discussed spin effect induces the electric dipole moment locally.

In Sec.~\ref{sec:general} we discuss general features of this spin effect.
Sec.~\ref{sec:instanton} is devoted to a  numerical demonstration of the effect
in a simple case of an instanton-like configuration.
In Sec.~\ref{sec:quenched} the quenched lattice simulations are used to show
that the quark electric dipole moment is induced both in
high- and low-temperature phases. Our conclusions are given in the last section, and the
appendix contains technical details of our simulations.

\section{General arguments}
\label{sec:general}

Quarks are the electrically charged spin-1/2 particles which carry certain magnetic dipole moments.
A strong enough external magnetic field
aligns quark spins and leads to appearance of a net magnetization.
A quantitative measure of the magnetization is given by the expectation value~\cite{ref:IS}
\begin{eqnarray}
\langle \bar\Psi \Sigma_{\alpha\beta} \Psi\rangle = \langle \bar\Psi\Psi\rangle\, \chi(F)\, q F_{\alpha\beta}\,,
\label{eq:chi:def}
\end{eqnarray}
where
\beqn
\Sigma_{\alpha\beta} = \frac{1}{2 i} [\gamma_\alpha \gamma_\beta - \gamma_\beta \gamma_\alpha]
\eeqn
is the relativistic spin operator, $F_{\alpha\beta} = \partial_\alpha a_\beta - \partial_\beta a_\alpha$
is the strength tensor of the external electromagnetic field, and $a_\alpha$ is the Abelian gauge field.
For the sake of simplicity we consider one quark
flavor, and omit the flavor and spinor indices in Eq.~\eq{eq:chi:def}.

The structure of the right hand side of Eq.~\eq{eq:chi:def} is obvious: the appearance of the
electromagnetic field strength tensor $F_{\alpha\beta}$ is dictated by a covariance argument,
while the appearance of the chiral condensate $\langle \bar\Psi \Psi\rangle$ is the natural
consequence of the dimensionality of the left and right parts of Eq.~\eq{eq:chi:def}. Moreover,
the left hand side of~\eq{eq:chi:def} violates the chiral symmetry similarly to the chiral condensate.
The coefficient of proportionality $\chi(F)$ is called the chiral magnetic susceptibility~\cite{ref:IS}.

There is another property which was used for the parametrization of the right hand side of Eq.~\eq{eq:chi:def}:
this equation is formulated in the $CP$-symmetric vacuum. If the vacuum is not $CP$-symmetric, then
Eq.~\eq{eq:chi:def} should be rewritten in a more general form:
\begin{eqnarray}
\langle \bar\Psi \Sigma_{\alpha\beta} \Psi\rangle_\theta = \langle \bar\Psi\Psi\rangle\, \chi(F)\, q F_{\alpha\beta}
+ \langle \bar\Psi\Psi\rangle\, {\widetilde \chi}(F)\, q {\widetilde F}_{\alpha\beta}\,,
\label{eq:chi:full}
\end{eqnarray}
where
\beqn
{\widetilde F}_{\alpha\beta} = \frac{1}{2} \varepsilon_{\alpha\beta\mu\nu} F_{\mu\nu}
\label{eq:Fdual}
\eeqn
is the dual field strength tensor. In Eq.~\eq{eq:chi:full} we used the subscript $\langle \cdots \rangle_\theta$
to stress that the spin expectation value is evaluated in a $CP$-odd environment, which can be induced, say, by a nonzero $\theta$-angle.
We omit the subscript $\theta$ in the right hand side of Eq.~\eq{eq:chi:full}. The new coefficient ${\widetilde \chi}$ is the electric
susceptibility of the vacuum which in general may be nonzero if $\theta \neq 0$.

One can rewrite Eq.~\eq{eq:chi:full} by introducing the dimensionless magnetic and electric dipole moments, respectively
\beqn
\mu_{\alpha\beta} = \chi(F)\, q F_{\alpha\beta}\,,
\qquad
\epsilon_{\alpha\beta} = {\widetilde \chi}(F)\, q {\widetilde F}_{\alpha\beta}\,,
\label{eq:mu:epsilon}
\eeqn
so that Eq.~\eq{eq:chi:full} is simplified:
\begin{eqnarray}
\langle \bar\Psi \Sigma_{\alpha\beta} \Psi\rangle_\theta = (\mu_{\alpha\beta} + \epsilon_{\alpha\beta})\, \langle \bar\Psi\Psi\rangle\,.
\label{eq:chi:full2}
\end{eqnarray}
Here the tensors $\mu_{\alpha\beta}$ and $\epsilon_{\alpha\beta}$ represent two orthogonal contributions,
\beqn
\mu_{\alpha\beta} \, \epsilon_{\alpha\beta} \equiv 0\,.
\eeqn

Note that in presence of a nonzero $\theta$-angle the tensor structure of the magnetization~\eq{eq:chi:full} is most general.
The nonzero $\theta$-angle leads to appearance of the pseudoscalar condensate
$\langle \bar\Psi\gamma_5\Psi\rangle$, the pseudotensor condensate $\langle \bar\Psi\gamma_5 t^a G^a_{\mu\nu}\Psi\rangle$,
where $G^a_{\mu\nu}$ is the gluon field strength tensor and $t^a$ are the generators of the gauge group, and other condensates.
The pseudoscalar condensate may give separate contributions to the chiral susceptibilities $\chi$ and ${\widetilde \chi}$ in
Eq.~\eq{eq:chi:full}, while it cannot mix the two terms in \eq{eq:chi:full} because of their mutual orthogonality.
As for the pseudotensor condensate, it may be split -- in terms of the Lorentz structures -- into two orthogonal terms that
are proportional to the electromagnetic field strength tensor $F_{\mu\nu}$ and its dual~\eq{eq:Fdual}, respectively. Therefore
the pseudotensor condensate may also contribute to the both terms of Eq.~\eq{eq:chi:full} leaving the general form of the magnetization~\eq{eq:chi:full} unchanged.

It is convenient to rewrite the magnetic and electric dipole moments~\eq{eq:mu:epsilon} via the expectation values~\eq{eq:chi:full}
in the vector form, respectively:
\begin{eqnarray}
\mu_i(q B)  & = & \frac{\langle \bar\Psi \gamma_5 \Sigma_{i0} \Psi\rangle}{\langle \bar\Psi\Psi\rangle}
\equiv \frac{1}{2} \varepsilon_{ijk} \frac{\langle \bar\Psi \Sigma_{jk} \Psi\rangle}{\langle \bar\Psi\Psi\rangle}\,,
\label{eq:mu}\\
\epsilon_i(q B)  & = & \frac{\langle \bar\Psi \Sigma_{i0} \Psi\rangle}{\langle \bar\Psi\Psi\rangle}
\quad \equiv \frac{1}{2} \varepsilon_{ijk} \frac{\langle \bar\Psi \gamma_5 \Sigma_{jk} \Psi\rangle}{\langle \bar\Psi\Psi\rangle}\,,
\label{eq:epsilon}
\end{eqnarray}
where we used the relation $\gamma_5 \Sigma_{\alpha\beta} = {\tilde{\Sigma}}_{\alpha\beta}$, and assumed that the external
electric field is absent, $F_{0i} = 0$. The interpretation of Eq.~\eq{eq:epsilon} as the electric dipole moment is obvious because
this vector quantity is defined by the electric component of the spin operator, $\Sigma_{i0}$.
Here spatial tensor indices are $i, j = 1, 2, 3$ and the Euclidean time is labeled by the index $0$.

In the real $CP$-invariant vacuum the electric dipole moment~\eq{eq:epsilon} should be identically equal to zero. However, the fluctuations
of the electric magnetic dipole moment may still be very strong. The real vacuum may be subdivided into topologically nontrivial $CP$-odd
domains in which the quarks may have possess an anomalous electric dipole moment along the direction of the magnetic field,
$\vec \epsilon \parallel \vec B$. In these domains the $CP$-odd electric dipole moment may be as strong as the magnetic dipole moment.
Below we check this conjecture using numerical simulations.

In order to characterize quantitatively the appearance of the quark electric dipole moment, we consider the following quantities:
the local magnetic dipole moment
\beqn
\sigma^M_i(x) = \frac{1}{2} \varepsilon_{ijk} \bar{\psi}\lr{x} \Sigma_{jk} \psi\lr{x}\,,
\label{eq:sigma:M}
\eeqn
the local electric dipole moment
\beqn
\sigma^E_i(x) = \bar{\psi}\lr{x} \Sigma_{i0} \psi\lr{x}\,,
\label{eq:sigma:E}
\eeqn
and the local chirality
\beqn
\rho_{5}\lr{x} = \bar{\psi}\lr{x} \gamma_{5} \psi\lr{x} \equiv \rho_L\lr{x} - \rho_R\lr{x}\,,
\label{eq:rho5}
\eeqn
which is the operator of the difference of the densities of left- and right- handed quarks.

\section{Instanton}
\label{sec:instanton}

Instanton is a topologically nontrivial solution to the classical equations of Yang--Mills theory.
Topological charge of the instanton is nonzero, and therefore it can be considered as a
simplest gauge filed configuration which may lead to the anomalous electric dipole moment.
\begin{figure}[htb]
  \includegraphics[width=4cm, angle=90]{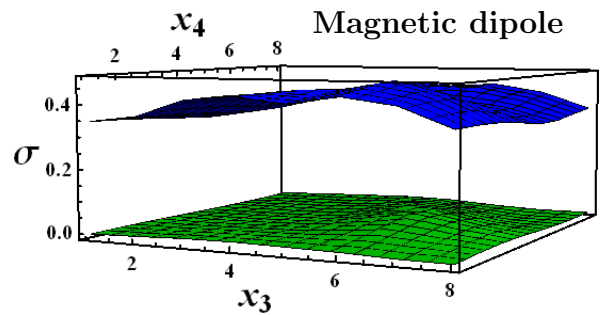}\\[-41mm]
  {\bf \large\hskip 35mm Magnetic dipole}\\[35mm]
  \caption{The local density of the magnetic dipole moment of quarks~\eq{eq:sigma:M} in the instanton background exposed to a strong magnetic field.
   The upper (blue) and lower (green) surfaces represent, respectively,
   the longitudinal, $\sigma^M_3(x)$, and a transverse, $\sigma^M_1(x)$,
   components of the dipole moment in a $34$-plane.}
  \label{fig:instanton:m}
\end{figure}
\begin{figure}[htb]
  \includegraphics[width=4cm, angle=90]{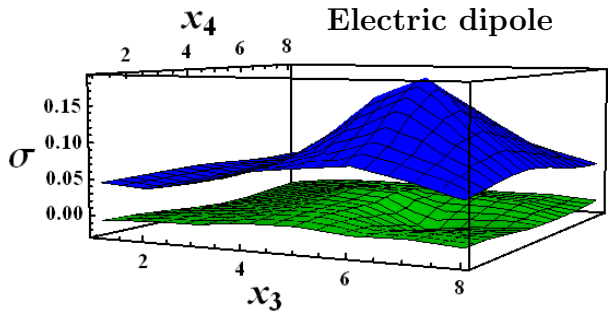}\\[-41mm]
  {\bf \large\hskip 35mm Electric dipole}\\[35mm]
  \caption{The same as in Fig.~\ref{fig:instanton:m} but for the local density of the electric dipole moment~\eq{eq:sigma:E}.}
  \label{fig:instanton:e}
\end{figure}

While a spectrum of the Dirac operator in the field of a single instanton is well-known~\cite{tHooft:76:1},
the presence of a uniform magnetic field makes the problem analytically intractable. Therefore we constructed an instantonlike smooth
configuration of the lattice gauge field with unit topological charge and numerically calculated fermionic propagators in such a
configuration (details of our numerical simulations are given in the appendix).
Next, we added the uniform magnetic field in the third direction, $B_i = B \delta_{i3}$ to this sample configuration.
We evaluated the local density $\sigma^M_i$ of the magnetic dipole moment~\eq{eq:sigma:M} in the background of this configuration.

Note that any gauge field configuration -- including the instanton -- has, in general, both zero and near-zero Dirac
eigenmodes. These classes of the Dirac modes have different physical meaning. For example,
the number of the zero modes is related to the topological charge of the
gauge field configuration according to the Atiyah-Singer theorem.
Consequently, the vacuum expectation value of the number of {\it zero} modes is proportional
to the susceptibility of the topological charge.

On the contrary, the Banks-Casher relation~\cite{Banks:80:1} states that
the vacuum expectation value of the density of the {\it near-zero} modes
is proportional to a different quantity, the chiral condensate. Therefore,
the near-zero modes carry information about the chiral properties
of the system, while the exact zero mode, in general, does not. In
Ref.~\cite{ref:magnetization} we have analytically related the chiral
magnetization to the spin structure of the near-zero modes in the external magnetic field
(we discuss this relation in the Appendix, and refer an interested reader to Ref.~\cite{ref:magnetization}
for more details). Thus, only near-zero modes were taken into account in our calculations of the chiral
magnetization.

We would like to make a remark that in various instanton gas and liquid models the individual zero modes of instantons
and antiinstantons do play an important role in the chiral properties of the system because in this interacting
system the zero-mode degeneracy is lifted out and the zero modes become eventually near-zero. In our illustrative
example we consider the single instanton like configuration so that zero mode does not contribute to the chiral
magnetization.

The density of the longitudinal ($i=3$) and the transverse ($i=1$) components of the magnetic dipole moment in the $34$-plane
of the background instantonlike configuration are plotted in Fig.~\ref{fig:instanton:m}. The distribution in the $12$-plane is similar to the one plotted in Fig.~\ref{fig:instanton:m}.
As one can expect, the magnetic dipole moment is predominantly directed along the magnetic field:
$|\sigma^M_3(x)| \gg |\sigma^M_{1,2}(x)|$ so that $\vec \sigma^M \parallel \vec B$ with a good accuracy. Notice that the spatial distribution of the density of the magnetic dipole moment is uniform similarly to the magnetic field.

The density of the electric dipole moment is shown in Fig.~\ref{fig:instanton:e}.
Similarly to the magnetic moment, the electric dipole moment~\eq{eq:sigma:E} is directed along the magnetic field:
$|\sigma^E_3(x)| \gg |\sigma^E_{1,2}(x)|$ so that $\vec \sigma^E \parallel \vec B$ with a good accuracy.
The magnitude of the electric dipole moment is smaller and, in contrast to the magnetic dipole moment, its distribution is peaked near the position of the instanton.

Summarizing this section, we observed the generation of the electric dipole moment in the background of the topologically nontrivial configuration of the smooth gauge fields. In the next section we study the same effect in a real ground state of non-Abelian gauge theory.

\section{Quenched QCD}
\label{sec:quenched}

\subsection{Fluctuations of electric dipole moment}

 The vacuum expectation value of the magnetic dipole moment of the quark~\eq{eq:sigma:M} is given by~\eq{eq:chi:def}:
\begin{eqnarray}
\langle \bar\Psi {\vec \sigma}^M \Psi\rangle = \langle \bar\Psi\Psi\rangle\, \chi(B)\, q {\vec B}\,.
\label{eq:chi:mu}
\end{eqnarray}
The chiral magnetic susceptibility $\chi$ was calculated numerically in Ref.~\cite{ref:magnetization}.

The average of the electric dipole moment~\eq{eq:sigma:E} is zero,
\beqn
\left\langle \bar\Psi {\vec \sigma}^E \Psi \right\rangle = 0\,,
\eeqn
since the density of topological charge changes its sign in different space-time domains, thus
supporting the global $CP$-invariance of the vacuum. However, locally the electric dipole moment ${\vec \sigma}^E$
may be nonzero, as we have seen in the previous section for the instanton case.

In order to measure the strength of the local fluctuations of the dipole moments we study the connected expectation values
\begin{eqnarray}
{\bigl\langle \bigl(\sigma^\ell_i\bigr)^2 \bigr\rangle}_{IR}
& = & {\Bigl\langle\bigl(\sigma^\ell_i - \bigl\langle\sigma^\ell_i\bigr\rangle\bigr)^2\Bigr\rangle}_{B, T}
\nonumber \\
& - & {\Bigl\langle \bigl(\sigma^\ell_i - \bigl\langle\sigma^\ell_i\bigr\rangle\bigr)^2\Bigr\rangle}_{B,T = 0}\,,
\label{ref:expectation}
\end{eqnarray}
where there is no summation over the indices $i$ and $\ell$.
In Eq.~\eq{ref:expectation} $\sigma^\ell_i$ is the $i$-th component of the magnetic ($\ell=M$) or electric ($\ell=E$)
dipole moment, and ${\langle\ldots\rangle}_{B, T}$ denotes the expectation value with respect to the thermal state at
temperature $T$ in the background magnetic field~$B$.
The subtraction of the expectation value at $B = 0, T = 0$ removes ultraviolet divergences and
yields physical results which are practically independent of the UV cutoff
(this question was discussed in Refs.~\cite{ref:CME,ref:magnetization} using the same set of the gauge field configurations).
The subscript $IR$ Eq.~\eq{ref:expectation} reflects the fact that the subtraction provides us
with the nonperturbative infrared (IR) value.

We calculated the fluctuations~\eq{ref:expectation} numerically in $SU\lr{2}$ lattice gauge theory with quenched massless chirally invariant quarks (details of simulations can be found in the appendix).
\begin{figure}
  \includegraphics[width=6cm, angle=-90]{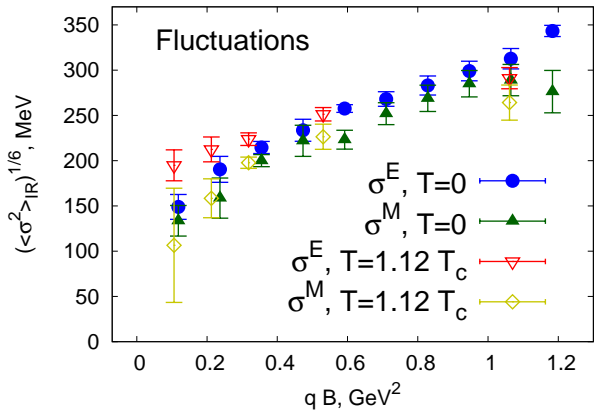}
  \caption{The fluctuations~\eq{ref:expectation} of the longitudinal (i.e., directed along the magnetic field) components of the magnetic~\eq{eq:sigma:M} and electric~\eq{eq:sigma:E} dipole densities vs the strength of the magnetic field in the confinement ($T = 0$) and in the deconfinement ($T = 1.12 \, T_c$) phases.}
  \label{fig:sigma2}
\end{figure}
In Fig.~\ref{fig:sigma2} we show the fluctuations~\eq{ref:expectation} of the longitudinal components of the electric dipole moment ($\cO = \sigma^E_3$, Eq.~\eq{eq:sigma:E}) both in the confinement phase, at $T=0$, and in the deconfinement phase, at $T=1.12\, T_c$. Here $T_c \approx 310$~MeV is the critical temperature of the confinement-deconfinement phase transition in the SU(2) gauge theory. For comparison, we also plotted the longitudinal component of the magnetic dipole moment [$\cO = \sigma^M_3$, Eq.~\eq{eq:sigma:M}]. One can see that the average squares of the magnetic moment and of the electric dipole moment are equal with a good precision. Both of them grow significantly with magnetic field. Taking into account the fact that the average electric dipole moment is zero, one can conclude that there are space-time domains within which the magnetic dipole moment and the electric dipole moment are either parallel or antiparallel, and are equally strong. It is interesting that the fluctuations of both magnetic and electric dipole moments are almost independent of the temperature even near the phase transition (the small mutual deviations in Fig.~\ref{fig:sigma2} are of the order of the error bars).

We also found that the fluctuations of the transverse components of the electric $\sigma^E_{1,2}$ and magnetic $\sigma^M_{1,2}$ dipole moments are almost independent of the magnetic field.  For these components the expectation value~\eq{ref:expectation} is compatible with zero.

Summarizing this subsection, we conclude that in the external magnetic field the quark develops the electric dipole moment, which is directed along the magnetic field. The fluctuations of the electric dipole moment are of the same order as the magnetic ones.

\subsection{Electric dipole moment and chirality}

In order to demonstrate that the electric dipole moment is closely related with the local chirality we calculate the correlator of the electric dipole moment with the chiral density~\eq{eq:rho5}:
\begin{eqnarray}
c\lr{ \rho_{5}, \sigma^{E,M}_i } = \frac{ \bigl\langle \rho_{5} \, \sigma^{E,M}_i \bigr\rangle}{\sqrt{\bigl\langle \rho_{5}^{2}\bigr\rangle} \sqrt{\bigl\langle\bigl(\sigma^{E,M}_i\bigr)^{2}\bigr\rangle }}\,.
\label{eq:correlation}
\end{eqnarray}
In this formula no summation over the index $i$ is implied.

\begin{figure}[htb]
  \includegraphics[width=6cm, angle=-90]{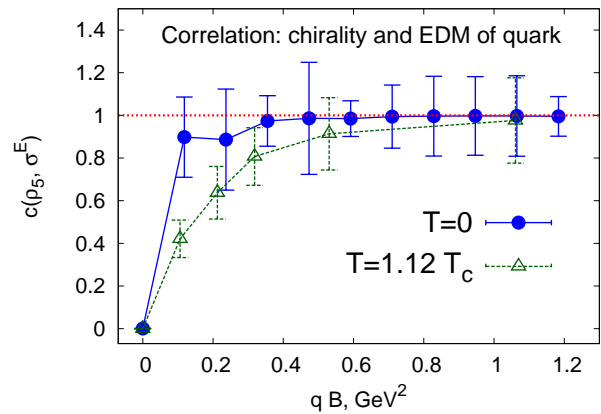}
  \caption{The correlation~\eq{eq:correlation} of the longitudinal component of the electric dipole moment~\eq{eq:sigma:E} of quark (EDM) with the chiral density~\eq{eq:rho5} vs the strength of the magnetic field in the confinement and deconfinement phases. The horizontal dotted line indicates the 100\% correlation.}
  \label{fig:correlation}
\end{figure}

We plot the correlator~\eq{eq:correlation} of the longitudinal electric dipole moment, $\sigma^E_3$, in Fig.~\ref{fig:correlation} both in the confinement phase and in the deconfinement phase. At zero temperature the correlator grows quickly and we observe almost the full (100\%) correlation of the quark electric dipole moment and the chirality even at weakest nonzero magnetic field. The strength of the effect is somewhat smaller in the deconfinement phase at weak magnetic fields due to thermal fluctuations. However, at strong magnetic fields, $q B \sim 1\,\mbox{GeV}^2$, the correlation function~\eq{eq:correlation} reaches the highest possible value, $c\lr{ \rho_{5}, \sigma^E_3} = 1$.

The transverse components $\sigma^E_{1,2}$ of the electric dipole moment~\eq{eq:sigma:E} are not locally correlated with the chirality~\eq{eq:rho5}. The same is true both for transverse $\sigma^M_{1,2}$ and longitudinal $\sigma^M_3$ components of the magnetic dipole moment of quark. Note that no correlation is observed in the absence of the external magnetic field.

Summarizing, at zero temperature the electric dipole moment is strongly correlated with the chiral density in the presence of the external magnetic field. The thermal fluctuations reduce this correlation, which is restored again as the strength of the magnetic field increases.

\section{Conclusion}

We demonstrated that the quark develops the anomalous electric dipole moment in the presence of a sufficiently strong external magnetic field. The anomalous moment is parallel to the direction of the magnetic field. The observed effect is a reflection of a topological structure of the QCD vacuum: the induced electric dipole moment is strongly correlated with the chiral density of quarks which, in turn, is correlated with the topological charge density. The sign of the electric dipole moment of the quark is a fluctuating quantity so that the electric dipole moment of the quark is zero on average, and no global $CP$-violation occurs. However, the local fluctuations of the longitudinal (i.e., parallel to the magnetic field) component of the quark electric dipole moment are as strong as the fluctuations of the conventional magnetic dipole moment.

The effect is observed at a single instanton gauge field configuration, as well as in a true vacuum state of a quenched non-Abelian gauge theory.
The effect is strong both in the confinement and deconfinement phases. The thermal fluctuations decrease the correlation of the longitudinal electric dipole moment with the chiral density at moderate magnetic fields. As the strength of the field increases, these two quantities become fully correlated.

\appendix
\section{Details of simulations}

The setup of our numerical simulations is the same as the one used in Refs.~\cite{ref:CME,ref:magnetization,ref:condensate}. We utilize lattice QCD with the simplest $SU(2)$ gauge group because the generation of the electric dipole moment originates in the chiral sector of QCD and therefore the number of colors is not crucial. In our simulations only valence quarks interact with the electromagnetic field. The effects of the virtual quarks on gluons are neglected because the inclusion of dynamical (sea) quarks makes the simulations computationally difficult, while the essential features of the studied effect remains intact in the quenched limit.

In Table~\ref{tbl:parameters} we present the parameters of our lattice simulations: the lattice geometry, $L_s^3 L_t$, the coupling constants $\beta$, the lattice spacings $a$, the spatial lattice extension $L = L_s a$, and the minimal value of the magnetic fields $ \sqrt{q B_\mathrm{min}}$.
\begin{table}[htb]
\begin{center}
\caption{Parameters of simulations.}
\label{tbl:parameters}
\begin{tabular}{ccccccc}
\hline
\hline
$T/T_c$ & $L_s$ & $L_t$ & $\beta$  & $a$, fm  & $L_s a$, fm  &  $ \sqrt{q B_\mathrm{min}}$, MeV \\
\hline
0     &  14    & 14   & 3.281   & 0.103     & 1.44 &  343  \\
1.12  &  16    &  6   & 3.325   & 0.095     & 1.33 &  371  \\
\hline
\hline
\end{tabular}
\end{center}
\end{table}

In order to implement chirally symmetric massless fermions on the lattice, we use Neuberger's overlap Dirac operator \cite{Neuberger:98:1}. Ultraviolet lattice artifacts are reduced with the help of the tadpole-improved Symanzik action for the gluon fields (see, e.g., Eq.~(1) in \cite{Luschevskaya:08:1}). The uniform magnetic field $B$ in the direction $\mu = 3$ is introduced into the Dirac operator by substituting $su\lr{2}$-valued vector potential $A_{\mu}$ with $u\lr{2}$-valued field: $A_{\mu}^{ij} \rightarrow A_{\mu}^{ij} + C_\mu \delta^{ij}$,
where $C_\mu  = B/2 \lr{x_{2} \delta_{\mu1} - x_{1} \delta_{\mu2}}$. Notice that both $su(2)$ and $u(1)$ are algebra-valued functions.

This expression is valid in the infinite volume. In order to make it consistent with the periodic boundary conditions in the spatial directions, we have introduced an additional $x$-dependent boundary twist for fermions on the lattice~\cite{Wiese:08:1}. The finiteness of the volume leads to the quantization of the total magnetic flux, so that
\begin{eqnarray}
\label{eq:bquant}
q B = 2 \pi \, k/L^{2}\,, \qquad k \in \mathbb{Z}\,,
\end{eqnarray}
where $q = 1/3\, e$ is the smallest (absolute value of) electric charge of the quark, and $L$ is the length of the lattice in the spatial direction.

In order to calculate the chiral expectation values~\eq{eq:chi:def}, \eq{ref:expectation} and \eq{eq:correlation} we use the  basis of the eigenmodes $\psi_{n}$ of the Dirac operator $\mathcal{D} = \gamma^{\mu} \, \lr{\partial_{\mu} - i A_{\mu}}$,
\begin{eqnarray}
\mathcal{D} \psi_{n} = \lambda_{n} \psi_{n}\,,\qquad \rho\lr{\lambda} = \vev{ \sum \limits_{n} \delta\lr{\lambda - \lambda_{n}} }\,,
\end{eqnarray}
where $\lambda_{n}$ are the eigenvalues of the Dirac operator.

The chiral condensate is calculated using the Banks-Casher formula \cite{Banks:80:1},
\begin{eqnarray}
\label{BanksCasher}
\left\langle \bar\Psi\Psi \right\rangle = - \lim \limits_{\lambda \rightarrow 0} \lim \limits_{V \rightarrow \infty} \, \frac{\pi \rho\lr{\lambda}}{V}
\end{eqnarray}
where $V$ is the total four-volume of Euclidean space-time, and $\rho\lr{\lambda}$ is the density of the Dirac eigenvalues. In the quenched approximation the averaging is performed over the gauge fields $A_{\mu}$ with the weight $\expa{ - S_{\mathrm{YM}}\lrs{A_{\mu}}}$, where $S_{\mathrm{YM}}\lrs{A_{\mu}}$ is Yang-Mills action. In the Euclidean space the spinor conjugation is given by the complex conjugation $\bar{\psi}_{\alpha} = \psi^{\dag}_{\alpha}$.

In \cite{ref:magnetization} we derived a magnetization analogue of the Banks-Casher formula~\eq{BanksCasher}:
\begin{eqnarray}
\langle \bar\Psi \Sigma_{\alpha\beta} \Psi\rangle  = - \lim_{\lambda \to 0}
\Bigl\langle\frac{\pi \nu(\lambda)}{V} \int d^4 x \, \psi^\dagger_\lambda(x) \,
\Sigma_{\alpha\beta}\, \psi_\lambda(x)\Bigr\rangle\,, \nonumber
\end{eqnarray}
which is used for evaluation of the expectation values containing the spin operator.
The expectation values involving four fermionic fields are evaluated using the following formula:
\begin{eqnarray}
\vev{\bar{\psi} O_{1} \psi \, \bar{\psi} O_{2} \psi}
& {=} & \tr\lr{\frac{1}{\mathcal{D} + m} \, O_{1}} \tr\lr{\frac{1}{\mathcal{D} + m} \, O_{2}}
\nonumber \\
& & - \tr\lr{\frac{1}{\mathcal{D} + m} \, O_{1} \, \frac{1}{\mathcal{D} + m} \, O_{2}},
\nonumber
\end{eqnarray}
where
$O_{1}$ and $O_{2}$ are some spinor operators. The result is then averaged over all configurations of the gauge fields. The Dirac propagator is evaluated by inverting the massive Dirac operator in the subspace spanned by $M$ Dirac eigenvectors which correspond to $M$ nonzero Dirac eigenvalues with smallest absolute values:
\begin{eqnarray}
\frac{1}{\mathcal{D} + m}\lr{x, y} = \sum \limits_{k < M} \frac{\bar{\psi}_{k}\lr{x} \, \psi_{k}\lr{y}}{\lambda_{k} + m}.
\end{eqnarray}
The value of $M$ is limited by the numerical procedure used to find the eigensystem of the Dirac operator (\texttt{ARPACK} in our case). We have used $M = 10$ and $M=12$ (after the subtraction of ultraviolet divergences the expectation values are almost independent of $M$ for $M \ge 10$, Ref.~\cite{ref:CME}).

\vskip -5mm
\begin{acknowledgments}
This work was partly supported by Grants RFBR 08-02-00661-a, and DFG-RFBR 436 RUS, by the grants for scientific schools No. NSh-679.2008.2 and No. NSh-4961.2008.2, by the Federal Program of the Russian Ministry of Industry, Science and Technology No. 40.052.1.1.1112, by the Russian Federal Agency for Nuclear Power, and by the STINT Institutional grant IG2004-2 025. P.V.B. is also supported by the personal grant of the Dynasty
foundation. The calculations were partially done on the MVS 50K at Moscow Joint Supercomputer Center.
\end{acknowledgments}

\end{document}